\documentclass[aps,preprint,prd,showpacs,nofootinbib]{revtex4}
\usepackage{amsmath}
\usepackage{graphicx}
\usepackage{subfigure}
\usepackage{bm}
\usepackage{amssymb}
\usepackage{mathtools}
\usepackage{enumerate}
\usepackage{color}
\usepackage[colorlinks,linkcolor=magenta,anchorcolor=cyan,citecolor=blue]{hyperref}

\newcommand{\bal}{\begin{aligned}}
\newcommand{\eal}{\end{aligned}}
\newcommand{\R}{\mathcal{R}}
\newcommand{\K}{\mathcal{K}}
\newcommand{\h}{\mathcal{H}}

\newcommand{\s}{\mathcal{S}}

\newcommand{\e}{\eta}

\begin{document}

\title{Implication of GW170817 for cosmological bounces}

\author{Gen Ye$^{1}$\footnote{yegen14@mails.ucas.ac.cn}}
\author{Yun-Song Piao$^{1,2}$\footnote{yspiao@ucas.ac.cn}}

\affiliation{$^1$ School of Physics, University of Chinese Academy of
    Sciences, Beijing 100049, China}

\affiliation{$^2$ Institute of Theoretical Physics, Chinese
    Academy of Sciences, P.O. Box 2735, Beijing 100190, China}

\begin{abstract}

The detection of GW170817 and its electromagnetic counterpart has
revealed the speed of gravitational waves coincides with the speed
of light, $c_T=1$. Inspired by the possibility that the physics
implied by GW170817 might be related with that for the primordial
universe, we construct the spatially flat stable (throughout the
whole evolution) nonsingular bounce models in the beyond Horndeski
theory with $c_T=1$ and in the degenerate higher-order
scalar-tensor (DHOST) theory with $c_T=1$, respectively. Though it
constricts the space of viable models, the constraint of $c_T=1$
makes the procedure of building models simpler.

\end{abstract}
\maketitle

\section{Introduction}

Inflation is a successful scenario of the early universe
\cite{Guth1981,Linde1982,Albrecht1982,Starobinsky1980}.
%It
%generates a nearly scale-invariant scalar spectrum, consistent
%with observation
%\cite{Planck2018:inflation,Planck2018:parameters}.
However, it is well-known that inflation suffered from the
singularity problem
\cite{Hawking1970:singularity,Guth2003:inflationary}. This
suggests that our understanding about the gravity and the early
universe is incomplete. Instead of going to the quantum regime and
studying the physics of the ``singularity", one might construct
classical nonsingular cosmological models alternative or
complementary to the inflation scenario. Bouncing cosmology is a
class of such models with different applications, see e.g.
\cite{Khoury2001:ekpyrotic,Piao:2003zm,Creminelli2006:starting}
for earlier studies,
\cite{Rubakov2014:null,Cai:2014bea,Battefeld:2014uga}
for recent reviews.

Building nonsingular cosmological models in the scalar field
theories has been still one of the endeavors. It had been observed
that the spatially flat bounce models constructed in Horndeski
theories \cite{Horndeski:1974wa} inevitably encounter
instabilities (or else the singularity in Lagrangian)
\cite{Rubakov2016:generalized,Kobayashi2016:generic}, the so-called
No-go Theorem, see also
Refs.\cite{Ijjas:2016tpn,Ijjas2017:fully,Dobre2018:unbraiding}
for the attempts in the Horndeski theory.
%\cite{Buchbinder2007:ekpyrotic,Koehn2015:non-singular}. It was
Recently, based on the effective field theory (EFT) of
cosmological perturbations, it has been found that
%scalar-tensor theories applied to cosmological scenarios
%\cite{Creminelli:2007effective,Piazza2013:effective,Langlois2013:essential,Langlois2017:effective,Cai2016:effective},
the solutions of fully stable (without ghosts, gradient instabilities, etc., throughout the whole
	evolution)
cosmological bounce do exist if one goes beyond Horndeski
\cite{Cai:2016thi,Creminelli:2016zwa}, see
Ref.\cite{Cai:2017dyi,Kolevatov:2017voe,Mironov:2018oec}
for the corresponding bounce models performed in full covariant
Lagrangians (of the beyond Horndeski theory
\cite{Gleyzes:2014dya}). The progress caused by "No-go" have also
stimulated lots of studies,
e.g.\cite{Kolevatov:2016ppi,Akama:2017jsa,Misonoh:2016btv,deRham:2017aoj,Yoshida:2017swb,Santoni:2018rrx,Boruah:2018pvq,Banerjee:2018svi}.

Beyond Horndeski theories are a subclass of the degenerate
higher-order scalar-tensor (DHOST) theory
\cite{Motohashi:2014opa,Langlois:2015cwa,Langlois:2017mxy,Langlois:2018dxi}.
Unlike in general relativity (GR), the propagating speed $c_T$ of
gravitational waves (GW) in the DHOST theory might deviate
considerably from the speed of light. Recently, the detection of
GW170817 \cite{gw170817} and its electromagnetic counterpart has
provided a precise measurement for the speed of GWs: it coincides
with the speed of light with deviations smaller than a few $\times
10^{-15}$, i.e. $c_T=1$. This measurement strictly constrained the
scalar-tensor theories responsible for the acceleration of the current
universe
\cite{Lombriser:2015sxa,Lombriser:2016yzn,Creminelli2017:dark,Sakstein:2017xjx,Ezquiaga2017:dark,Baker:2017hug,Langlois:2017dyl,Boran:2017rdn}.
Though the physics implied by GW170817 seems not straightly
related with that for the primordial universe, undeniably, such
potential relevance will be interesting.

In this paper, inspired by the implication of GW170817, we will
construct the stable cosmological bounce models with $c_T=1$.
Using the ADM metric, we replace the covariant $c_T=1$ DHOST
Lagrangian with its ADM form (Sect.\ref{sec:lagrangian}), and
perform the perturbation calculations with it. We construct fully
stable bounce models in the beyond Horndeski theory with $c_T=1$
(Sect.\ref{sec:bh model}), which is a special subclass of the DHOST theory,
and in the full $c_T=1$ DHOST theory (Sect.\ref{sec:dhost model}),
respectively.

\section{DHOST theory with $c_T=1$}\label{sec:lagrangian}

\subsection{The Lagrangians}

We begin with the covariant Lagrangian of the beyond Horndeski theory
with $c_T=1$ \cite{Creminelli2017:dark}
\begin{equation}\label{ct1_L_bh_cov}
\bal
\mathcal{L}_{c_T=1}^{bH}=\sqrt{-g}L_{c_T=1}^{bH}=
\sqrt{-g}\Big[&G_2(\phi,X)+G_3(\phi,X)\square\phi+B_4(\phi,X){R}\\
&-\frac{4}{X}B_{4,X}(\phi,X)(\phi^\mu\phi^\nu\phi_{\mu\nu}\square\phi-\phi^\mu\phi_{\mu\nu}\phi_\lambda\phi^{\lambda\nu})\Big],
\eal
\end{equation}
where $\nabla_\mu\phi\equiv \phi_\mu$,
$\nabla_\nu\nabla_\mu\phi\equiv\phi_{\mu\nu}$ and
$X\equiv\phi_\mu\phi^\mu$.

We adopt the ADM metric
\begin{equation}\label{adm metric}
ds^2=-N^2dt^2+h_{ij}(dx^i+N^idt)(dx^j+N^jdt),
\end{equation}
where $N$ is the lapse, $N_i$ is the shift, $h_{ij}$ is the
spatial metric. We will use $\e=\phi$ as the time coordinate in the
FRW metric,
%\begin{equation}\label{FRW metric}
$ds^2=-N(\e)^2d\e^2+a^2|d\vec{x}|^2$.
%\end{equation}
Dynamics of $\phi$ has been absorbed into $N(\e)$, since
$\phi'\equiv{d\phi}/{d\e}=1$.

In the unitary gauge $\delta\phi=0$, the covariant Lagrangian
$L_{c_T=1}^{bH}$ \eqref{ct1_L_bh_cov} may be rewritten in the ADM
form
%is
%equivalent to the ADM Lagrangian
\cite{Gleyzes:2014dya},
%which is,
%based on the constant $\e$ hypersuface,
\begin{equation}\label{ct1_L_bh_adm}
L^{bH}_{c_T=1}=P(N,\e)+Q(N,\e)K+A(N,\e)(\R-\K_2),
\end{equation}
where $\R\equiv h^{ij}R_{ij}$ is the Ricci scalar on the spacelike
hypersurface, $K\equiv h^{ij}K_{ij}$ is the extrinsic curvature on
the spacelike hypersurface and $\K_2\equiv K^2-K_{ij}K^{ij}$. The
coefficients $P(N,\e)$, $Q(N,\e)$ and $A(N,\e)$ are related with
$G_2$, $G_3$ and $B_4$ in \eqref{ct1_L_bh_cov} by
\begin{equation}\label{PQA}\bal
P(X,\phi)=G_2-&\sqrt{-X}\int\frac{G_{3,\phi}}{2\sqrt{-X}}dX,\\
Q(X,\phi)=-\int G_{3,X}\sqrt{-X}dX&+2(-X)^{3/2}\int\frac{XB_{4,X\phi}-B_{4,\phi}}{X^2}dX,\\
A(X,\phi)&=B_4. \eal
\end{equation}

The covariant Lagrangian $L^{DHOST}_{c_T=1}$ of the DHOST theory with
$c_T=1$ has been identified in Ref.\cite{Langlois:2017dyl}. As
pointed out in Ref.\cite{Creminelli2017:dark}, $L^{DHOST}_{c_T=1}$
may be obtained by performing a conformal rescaling $g_{\mu\nu}\to
C(\phi,X)g_{\mu\nu}$ to $L^{bH}_{c_T=1}$. Since the light cone is
not altered, the corresponding DHOST theory will maintain $c_T=1$.
Therefore, the ADM Lagrangian of DHOST theories with $c_T=1$ may be
straightly calculated by rescaling
\[N\to\sqrt{C}N\qquad h_{ij}\to Ch_{ij}\qquad h^{ij}\to C^{-1}h^{ij}\]
where $C=C(N,\e)$. %We need the following formula during the calculation
%\begin{equation}\label{k_ij}
Here, without loss of generality, we will set $N_i=0$ in the
calculation. Considering $K_{ij}=\frac{1}{2N}(h_{ij}'-\nabla_i
N_j-\nabla_j N_i)$,
%\end{equation}
%Without loss of generality, we can set the shift functions $N_i=0$ in the calculation here.
after some integrations by parts and redefinition of
coefficients, we have
\begin{equation}\label{ct1_DHOST_L_adm}
\bal\mathcal{L}^{DHOST}_{c_T=1}=N\sqrt{h}L^{DHOST}_{c_T=1}=N\sqrt{h}&\left[P+QK+A(\R-\K_2)-\frac{3AB^2}{2N^2}N'^2-\frac{2AB}{N}N'K\right.\\&+\left.\frac{B}{a^2}\left(2\frac{A}{N}+2A_N-\frac{AB}{2}\right)(\partial
N)^2\right],\eal
\end{equation}
where $B(N,\e)=\partial_N(\log C)$. It can be checked that this
ADM Lagrangian is equivalent to the covariant $L^{DHOST}_{c_T=1}$
showed in Ref.\cite{Langlois:2017dyl}. When $B=0$ (or $C=const.$),
${L}^{DHOST}_{c_T=1}$ reduces to the beyond Horndeski Lagrangian
$L^{bH}_{c_T=1}$ in \eqref{ct1_L_bh_adm}. In order to write out
\eqref{ct1_DHOST_L_adm}, we have absorbed the term linear in $N'$
into $P+QK$ by
\[\frac{f(N,\eta)}{N}N'=n^{\mu}\nabla_\mu F-\frac{1}{N} \partial_\eta F=-F K-\frac{1}{N}\partial_\eta F\]
where $F\equiv\int f dN$.

\subsection{The EFT of scalar perturbation}

The quadratic order EFT of the DHOST theory is \cite{Langlois:2017mxy}
\begin{equation}\label{EFT}
\begin{split}
& S^{quad} = \int d^3x \,  d\eta \,  a^3  \frac{M^2}2\bigg\{ \delta K_{ij }\delta K^{ij}- \left(1+\frac23\alpha_L\right)\delta K^2  +(1+\alpha_T) \bigg( \R \frac{\delta \sqrt{h}}{a^3} + \delta_2 R \bigg)\\
&  + \h^2\alpha_K \delta N^2+4 \h \alpha_B \delta K \delta N+ ({1+\alpha_H}) \R  \delta N   +  4 \beta_1  \delta K  \delta N'   + \beta_2  \delta N'^2 +  \frac{\beta_3}{a^2}(\partial_i \delta N )^2
\bigg\}
\end{split}
\end{equation}
%with degenerate conditions\footnote{In \cite{Langlois2017:effective}, there are two separate sets of degenerate conditions. We only write down the one satisfied by theory \eqref{ct1_DHOST_L_adm}.}
Contracting \eqref{ct1_DHOST_L_adm} with \eqref{EFT}, we can
directly read off the effective coefficients in EFT \eqref{EFT},
\begin{equation}
\begin{aligned}
\frac{M^2}{2}= NA,\qquad \alpha_L &=0,\qquad \alpha_T=0,\\
\frac{M^2}{2}\h^2\alpha_K=L_N+\frac{1}{2}NL_{NN},\qquad &\frac{M^2}{2}4\h\alpha_B=NL_{NK}+2\h L_{N\s},\\
\frac{M^2}{2}({1+\alpha_H})=&A+NA_N,\\
\frac{M^2}{2}4\beta_1=-2AB,\qquad\frac{M^2}{2}\beta_2=-\frac{3AB^2}{2N}&,\qquad\frac{M^2}{2}\beta_3=NB\left(2\frac{A}{N}+2A_N-\frac{AB}{2}\right),
\end{aligned}
\end{equation}
where $\mathcal{H}/N\equiv \frac{da/d\e}{aN}=H$, and
$L^{DHOST}_{c_T=1}=L$ is set for simplicity. Degenerate
conditions have been checked
\begin{equation}\label{degenerate1}
\alpha_L= 0,\qquad\beta_2=-6\beta_1^2,
\end{equation}
\begin{equation}\label{degenerate2}
\beta_3=
-2N\beta_1\left[2(1+\alpha_H)+N\beta_1(1+\alpha_T)\right].
\end{equation}
Compared with that in Ref.\cite{Langlois:2017mxy}, the condition
\eqref{degenerate2} has been slightly modified, since we have not
necessarily $N(\e)=1$ here.

Use the scalar perturbation \footnote{When $N_i\ne0$, $N'$ in
$\mathcal{L}^{DHOST}_{c_T=1}$ \eqref{ct1_DHOST_L_adm} should be
promoted to $N'-N^i\partial_i N$.}
\begin{equation}\label{pert}
N_i\equiv\partial_i\psi,\qquad h_{ij}\equiv
a^2e^{2\zeta}\delta_{ij}
\end{equation}
%Notice we no longer assume $N_i=0$ here.
to expand \eqref{ct1_DHOST_L_adm} or EFT \eqref{EFT}. In the
corresponding result, $\delta N'\zeta'$ is absorbed into
$\tilde{\zeta}'^2$ by replacing $\zeta$ with a new variable
$\tilde{\zeta}=\tilde{\zeta}(\zeta,\delta N)$. Using $\delta
L/\delta\psi=0$, and after some integrations by parts, we get the
quadratic order Lagrangian of $\zeta$,
\begin{equation}\label{quad L}
\mathcal{L}_2=a^3\frac{M^2}{2}\left[U\zeta'^2-V\frac{(\partial\zeta)^2}{a^2}\right]
\end{equation}
with coefficients
%\begin{equation}\label{u}
%U  = \frac{1}{(1/N+N\alpha_B-(N\beta_1)' /\h )^2}  \bigg[\alpha_K+ 6\alpha_B^2 - \frac{6}{a^3 \h^2 M^2} \frac{d}{d\eta} \left( a^3 \h  M^2 \alpha_B \beta_1 \right) \bigg]
%\end{equation}
\begin{equation}\label{u}
U=\frac{\Sigma}{\gamma^2}+\frac{6}{N^2},
\end{equation}
%\begin{equation}\label{v}
%V =  \frac{2N}{a M^2 }\frac{d}{d\eta}\bigg[\frac{a M^2 \big( (1+\alpha_H)/N+\beta_1(1+\alpha_T)\big)}{\h (1/N+N\alpha_B)-(N\beta_1)'}\bigg] -2 (1+\alpha_T)
%\end{equation}
\begin{equation}\label{v}
V=2\left[\frac{N}{a M^2 }\frac{d}{d\eta}\left(a\mathcal{M}\right)-
1\right],
\end{equation}
where \footnote{The $\gamma$ in \eqref{gamma} is related to the $\gamma$
in Refs.\cite{Ijjas2018:space-time,Ijjas:2016tpn} by
$2A\gamma\to \gamma$.}
\begin{equation}\label{gamma}
\gamma\equiv
\frac{\mathcal{H}}{N}+N\mathcal{H}\alpha_B-(N\beta_1)',
\end{equation}
\begin{equation}\label{sigma}
\Sigma\equiv\mathcal{H}^2\left[\alpha_K+6\left(\alpha_B^2-\frac{\gamma^2}{\mathcal{H}^2N^2}\right)-18\alpha_B\beta_1-\frac{6(\mathcal{H}M^2\alpha_B\beta_1)'}{\mathcal{H}^2M^2}\right],
\end{equation}
\begin{equation}\label{m}
\mathcal{M}\equiv\frac{M^2}{\gamma}\left[(1+\alpha_H)/N+\beta_1\right].
\end{equation}
%We have defined $\Sigma$ in such a way that the conventional GR term  $6/N^2$ in $U$ is singled out.
The absence of ghost suggests \[U>0.\] The sound speed of scalar
perturbation is
\[c_S^2=V/U.\]
Gradient stability suggests $c^2_S>0$.

\section{Stable bounce models}

We will construct the fully stable (pathology-free) bounce models
in the beyond Horndeski theory \eqref{ct1_L_bh_adm} and DHOST theory
\eqref{ct1_DHOST_L_adm} with $c_T=1$, respectively. Both actually
belong to the subclasses of the full DHOST theory. We will follow the
method in Refs.\cite{Kolevatov:2017voe,Mironov:2018oec}.

\begin{figure}
   \centering
    \includegraphics[width=2.5in]{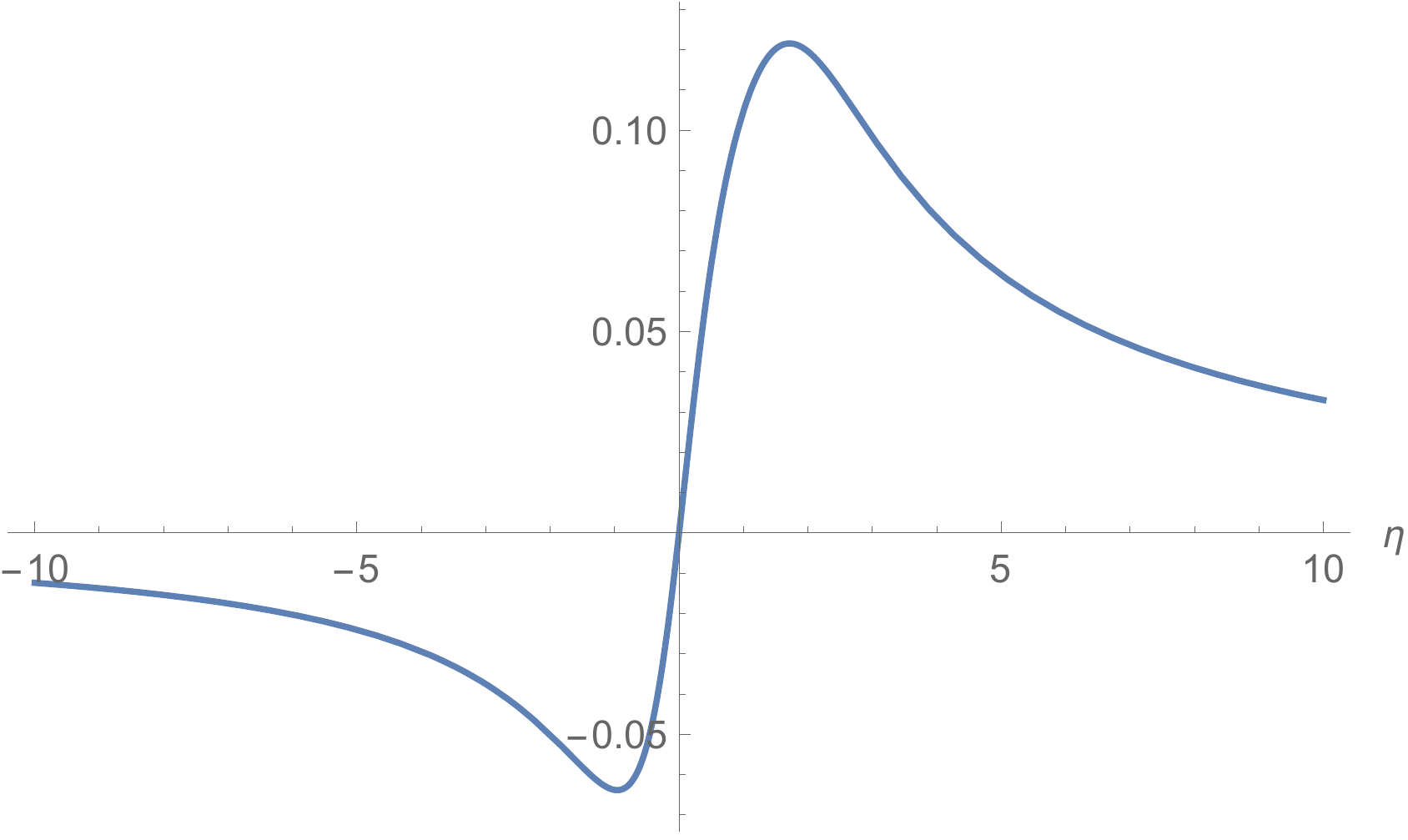}
    \caption{The evolution of $H$ with $p_i=8$ and $p_f=3$.}\label{H-model}
\end{figure}

We first set the evolutions of background (the Hubble parameter
$H$ and $N$). In our model, $H={\mathcal{H}}/{N}$ follows
\begin{equation}\label{H bh}
{\mathcal{H}}/{N}=\frac{\e}{p(\e)(1+\e^2)}
\end{equation}
with
\begin{equation}\label{p}
p(\e)=p_i+\frac{1+\tanh\left(\frac{\e-\e_p}{\tau_p}\right)}{2}(p_f-p_i),
\end{equation}
where $p_f$, $p_i$, $\e_p$ and $\tau_p=const$. Initially $\e\ll
-1$, ${\cal H}<0$, the universe contracts with $p(\e)=p_i$
($p_i\gg 1$ corresponds to the ekpyrotic contraction
\cite{Khoury2001:ekpyrotic}). Cosmological bounce occurs at
$\e=0$. Hereafter, the universe expands, and ${\cal H}>0$ has the
desired asymptotic form $\sim 1/(p_f\e)$, see Fig.\ref{H-model}.
Meanwhile $N$ follows
\begin{equation}\label{n}
x(\e)\equiv
\frac{1}{N}=x_i+\frac{1+\tanh\left(\frac{\e-\e_x}{\tau_x}\right)}{2}(x_f-x_i),
\end{equation}
where $x_f, x_i, \e_x$ and $\tau_x=const$. The choice of
Refs.\cite{Kolevatov:2017voe,Mironov:2018oec} is equivalent to setting
$p_i=p_f=3$ and $x_i=x_f=1$ (equivalently ${\dot \phi}=1$) in
(\ref{p}) and (\ref{n}), respectively.

\subsection{In beyond Horndeski theory}\label{sec:bh model}

%We will construct the fully stable bounce model in beyond
%Horndeski theory.
We set $M_p^2=(8\pi G)^{-1}=1$, and write $P(N,\e),Q(N,\e)$ and
$A(N,\e)$ in \eqref{ct1_L_bh_adm} as
\begin{equation}\label{free functions bh}
\bal P(N,\e)&=g_1(\e)\frac{1}{2N^2}+g_2(\e)\frac{1}{N^4}+g_3(\e),\\
Q(N,\e)&=0,\\
A(N,\e)&=\frac{1}{2}+f_1(\e)\frac{1}{N^2},\eal
\end{equation}
%The constant part of $A(N,\e)$ corresponds to Einstein gravity.
where the $N$-dependent part of $A(N,\e)$ sets the
coefficient $\sim B_{4,X}(\phi, X)\neq 0$ in $L_{c_T=1}^{bH}$
\eqref{ct1_L_bh_cov}, and is required for the fully stable bounce
\cite{Cai:2016thi,Creminelli:2016zwa,Cai:2017dyi,Kolevatov:2017voe}.
$Q(N,\e)$ is related with the cubic Galileon $G_3(\phi,X)\Box\phi$
in $L_{c_T=1}^{bH}$ \eqref{ct1_L_bh_cov}, see
\cite{Qiu:2011cy,Easson:2011zy,Cai:2012va,Qiu:2013eoa} for the
so-called G-bounce and \cite{Koehn:2013upa} for
super-bounce. However, $G_3(\phi,X)\Box\phi$ only moves the
period of $c_S^2<0$ to the outside of the bounce phase, but cannot
dispel it completely, as pointed out in
Refs.\cite{Ijjas:2016tpn,Easson:2011zy}. Thus we set
$Q(N,\e)=0$ for simplicity.

Since $Q(N,\e)=0$, \eqref{gamma} is simplified as
\begin{equation}
\label{gammabH}\gamma=\frac{\mathcal{H}}{AN}(A-NA_N),
\end{equation}
noting $\beta_1=0$ in the beyond Horndeski theory \eqref{ct1_L_bh_cov}. %$\gamma$ appears in the denominators of \eqref{u} and \eqref{v}.
To avoid possible divergence of $U$ induced by $\gamma=0$ (usually
called $\gamma$-crossing
\cite{Ijjas:2016tpn,Ijjas2017:fully,Ijjas2018:space-time}), we
choose $\Sigma$ in Eq.(\ref{sigma}) as
\begin{equation}\label{nsgl u}
\Sigma=c_1(\e)\gamma^2.
\end{equation}
$U>0$ (avoiding the ghost instability) can be insured by adjusting
$c_1(\e)$. $\gamma$-crossing will bring a singularity in unitary
gauge \cite{Ijjas2018:space-time}. However, as pointed out in
Ref.\cite{Mironov:2018oec}, this singularity does not affect the
proof of the No-go Theorem
\cite{Rubakov2016:generalized,Kobayashi2016:generic}.

According to \eqref{ct1_L_bh_adm}, we have the equations of ${\cal
H}$ and $N$ as follows,
\begin{equation}\label{eom bh}
\bal
6\frac{\h^2}{N^2}(A-NA_N)&=-P-NP_N-3\frac{\h}{N} (NQ_N),\\
\frac{4}{aN}\left(\frac{a'}{N}A\right)'&=-P-2\frac{\h^2}{N^2}A+\frac{1}{N}(Q_{\e}+Q_NN').
\eal
\end{equation}
One can solve out $g_1(\e)$, $g_2(\e)$ and $g_3(\e)$ in $P(N,\e)$
algebraically by considering Eqs.\eqref{nsgl u} and \eqref{eom
bh}, which are showed in Appendix \ref{apdx:bH}.

Substituting the corresponding solutions into \eqref{m}, we have
\[\mathcal{M}=\frac{1-4f_1^2x^4}{2\mathcal{H}(1+6f_1x^2)}.\]
We choose $f_1(\e)$ as
\begin{equation}\label{f1}
f_1(\e)=c_2(\e)\frac{c_3(\e)\mathcal{H}(\e)+1}{2x^2(\e)},\qquad
c_2(0)=1,
\end{equation}
%where $c_2(\e)$ and $c_3(\e)$ are free functions. In the future
%limit,
to make $\mathcal{M}$ not divergent at ${\cal H}=0$. $V>0$
(avoiding the gradient instability) can be insured by adjusting
$c_2(\e)$ and $c_3(\e)$, noting $1-4f_1^2x^4=0$ at
${\cal H}=0$.

\begin{figure}
    \centering
    \subfigure[$g_1(\e)$]{\includegraphics[width=2.5in]{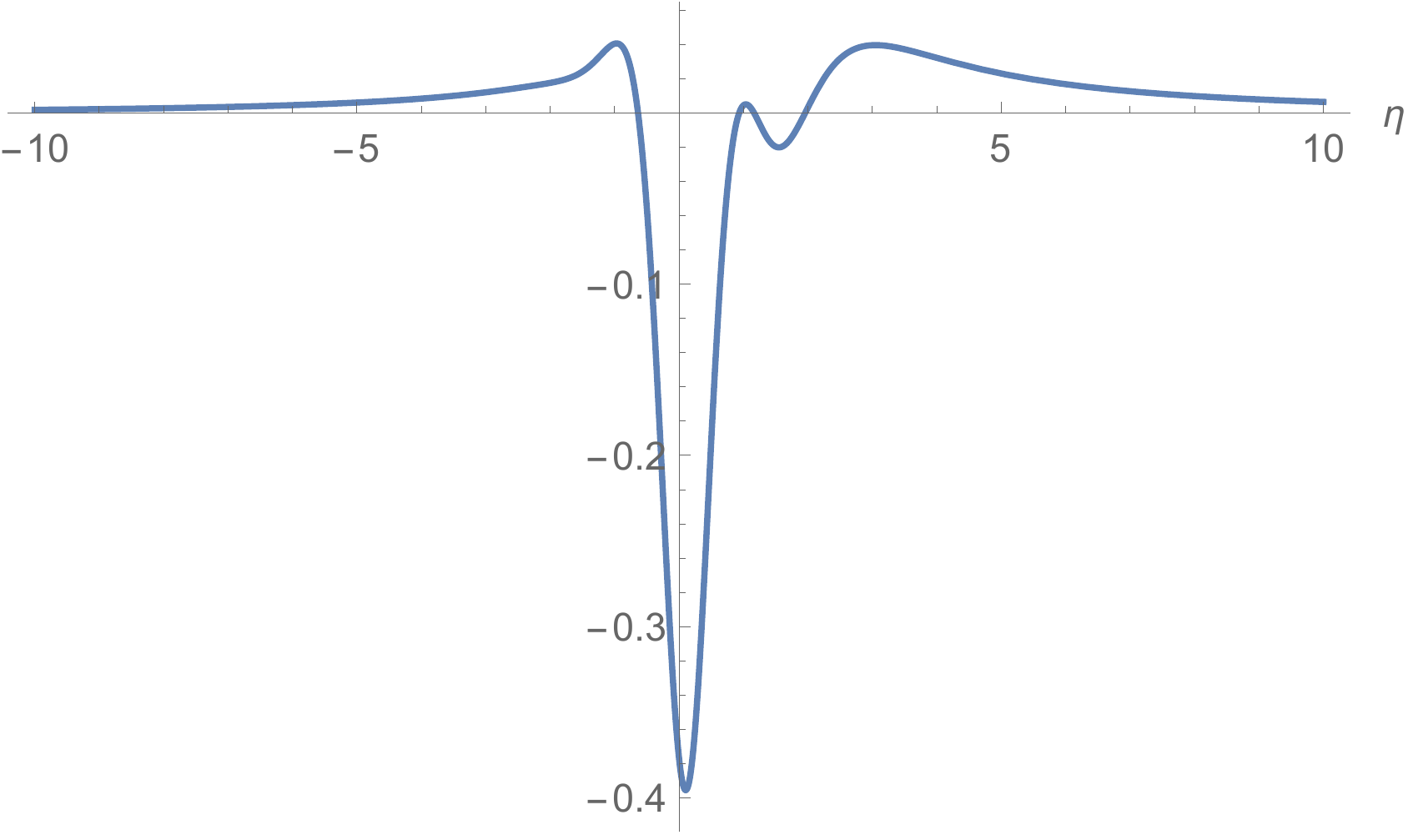}}
    \subfigure[$g_2(\e)$]{\includegraphics[width=2.5in]{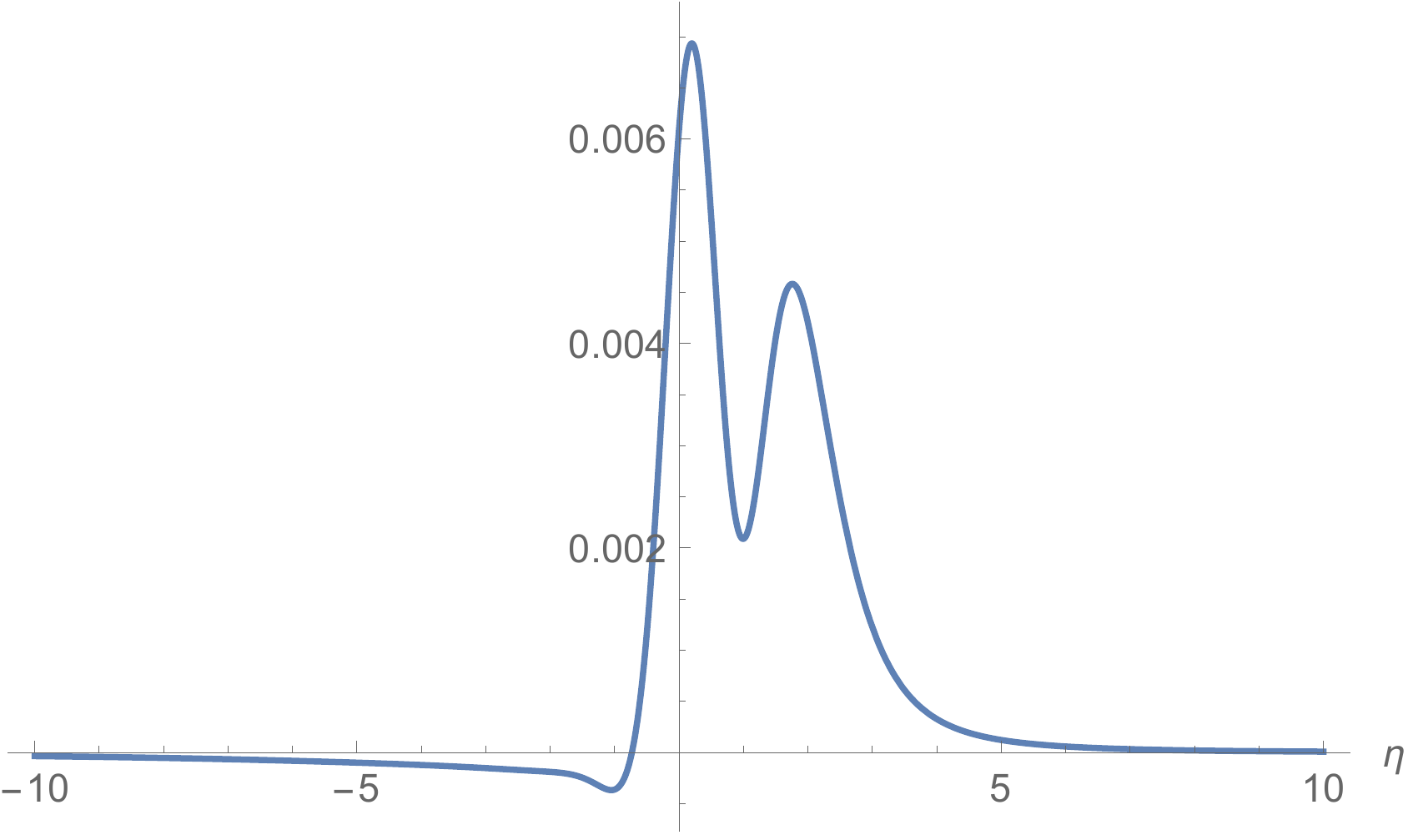}}
    \subfigure[$g_3(\e)$]{\includegraphics[width=2.5in]{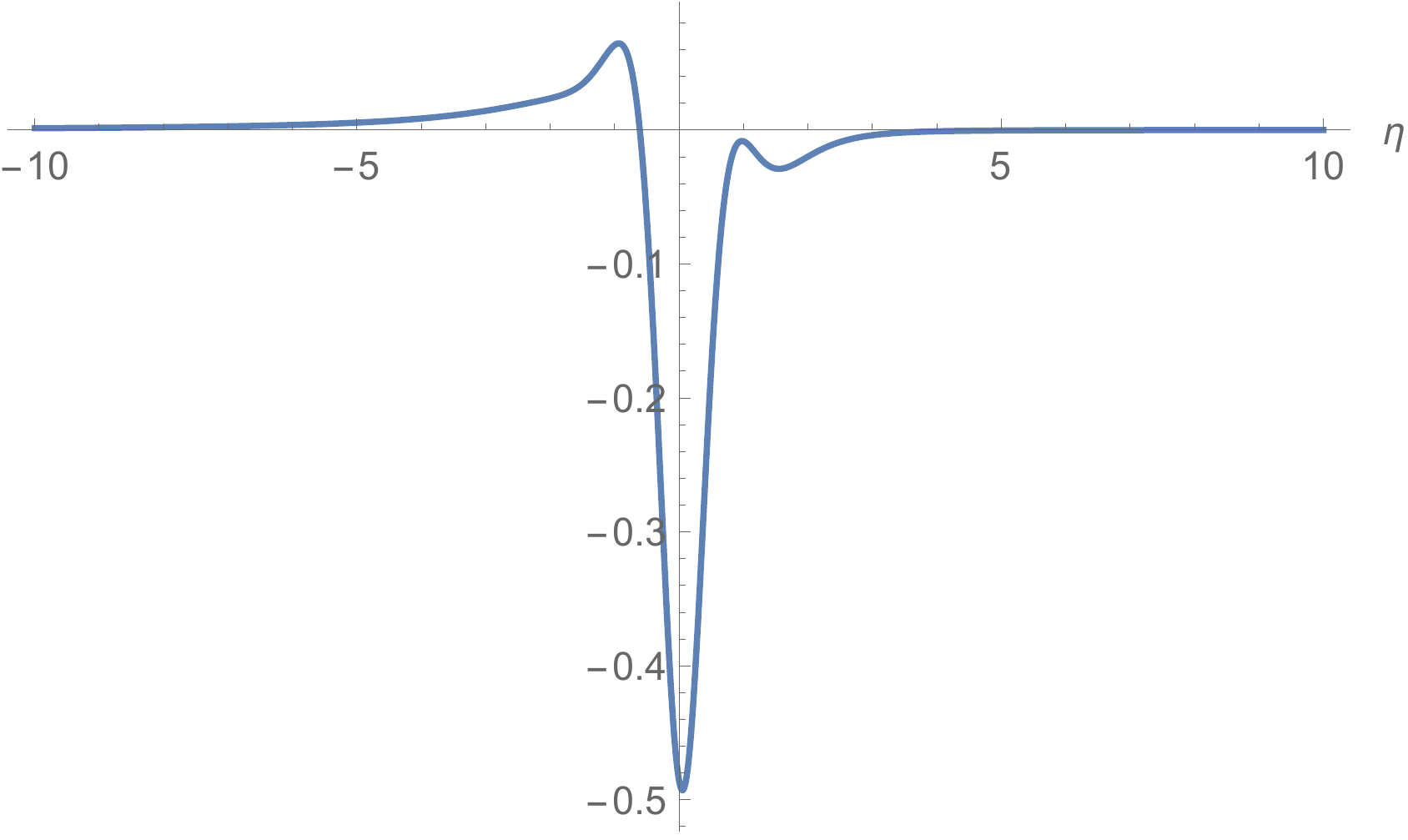}}
  \subfigure[$f_1(\e)$]{\includegraphics[width=2.5in]{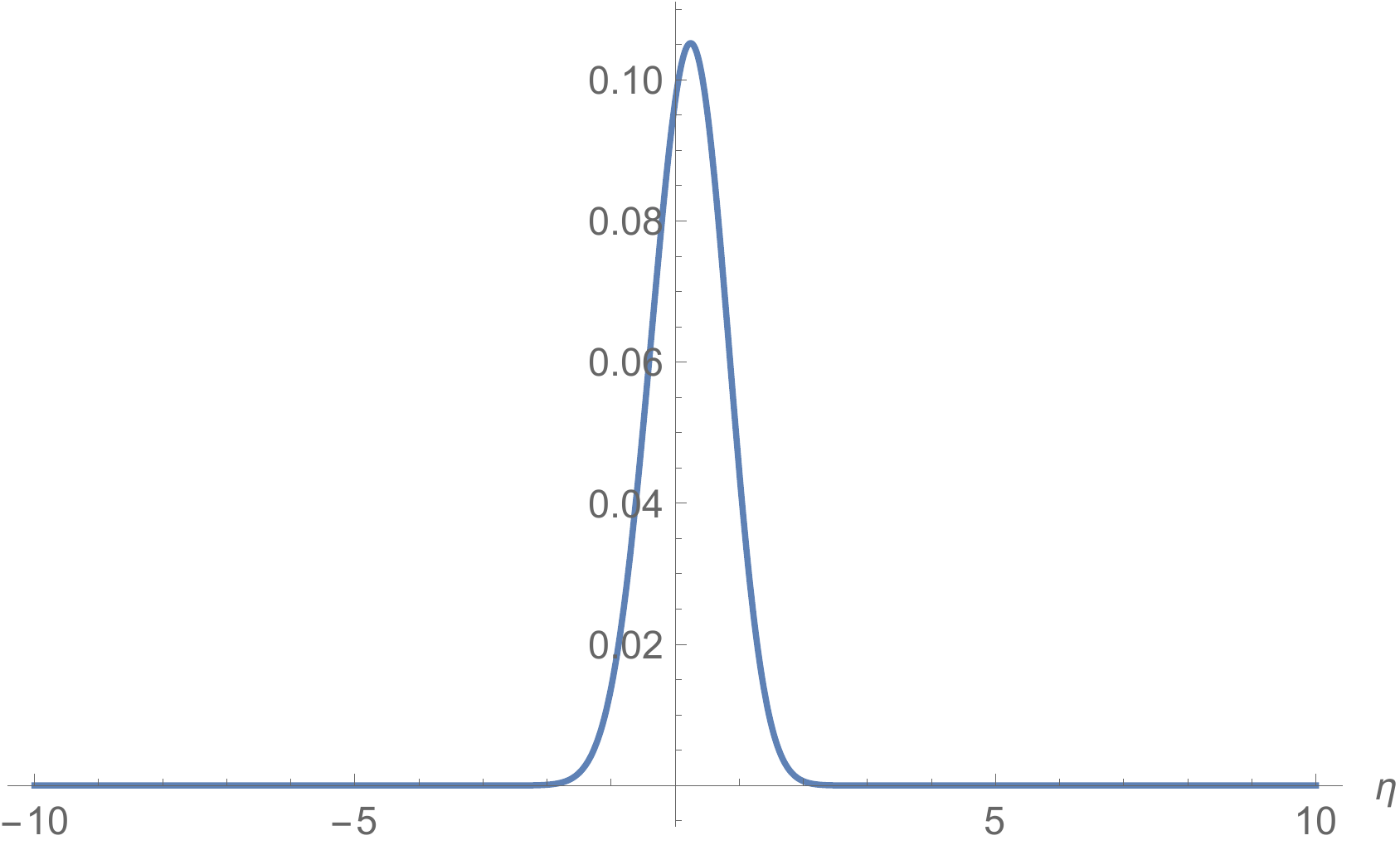}}
\caption{Coefficients of the beyond Horndeski
Lagrangian \eqref{free functions bh} in our bounce
model.}\label{bh-model-L}
\end{figure}

\begin{figure}
    \centering
   \subfigure[$U$]{\includegraphics[width=2.5in]{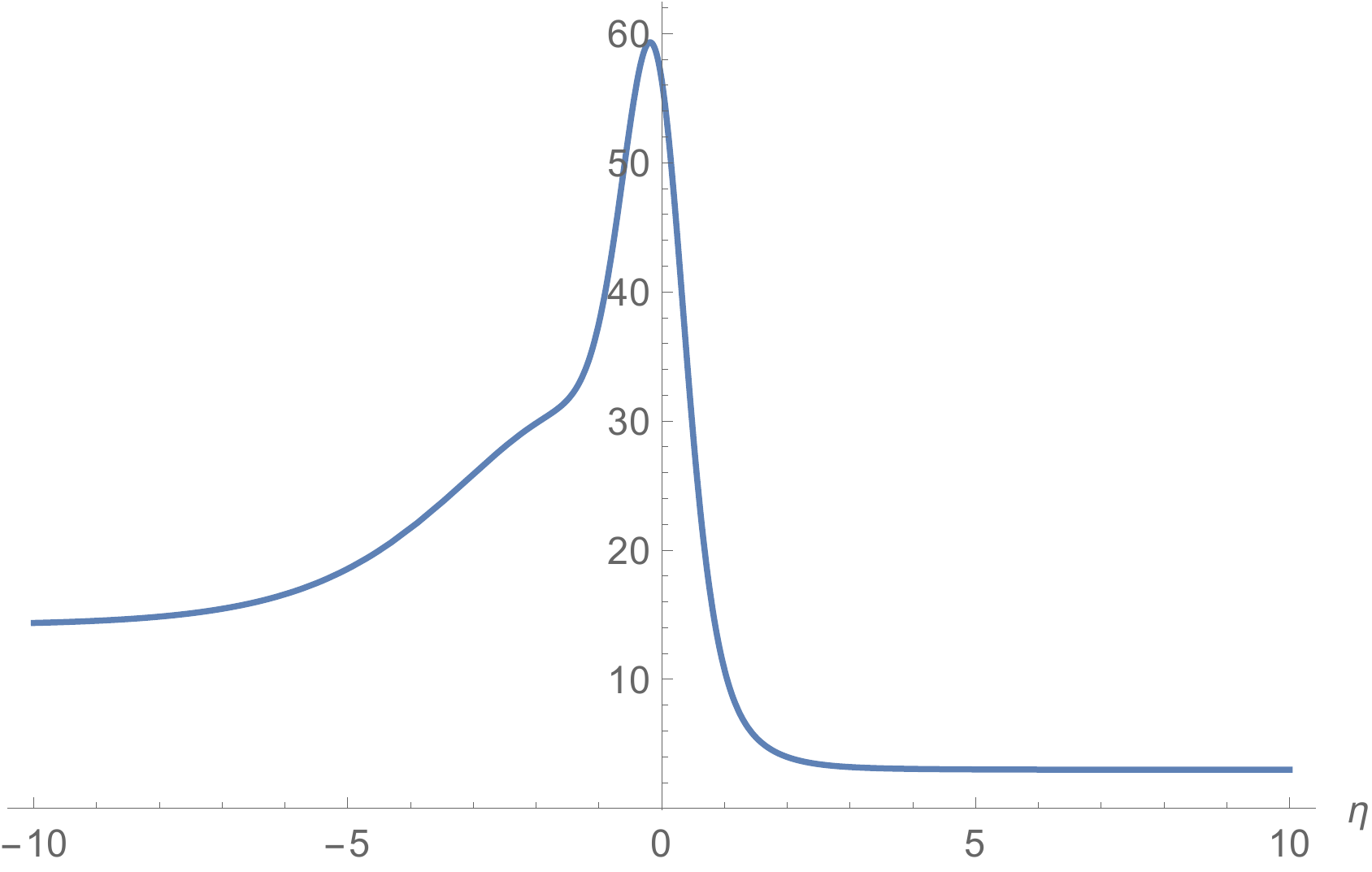}}
    \subfigure[$c_S^2$]{\includegraphics[width=2.5in]{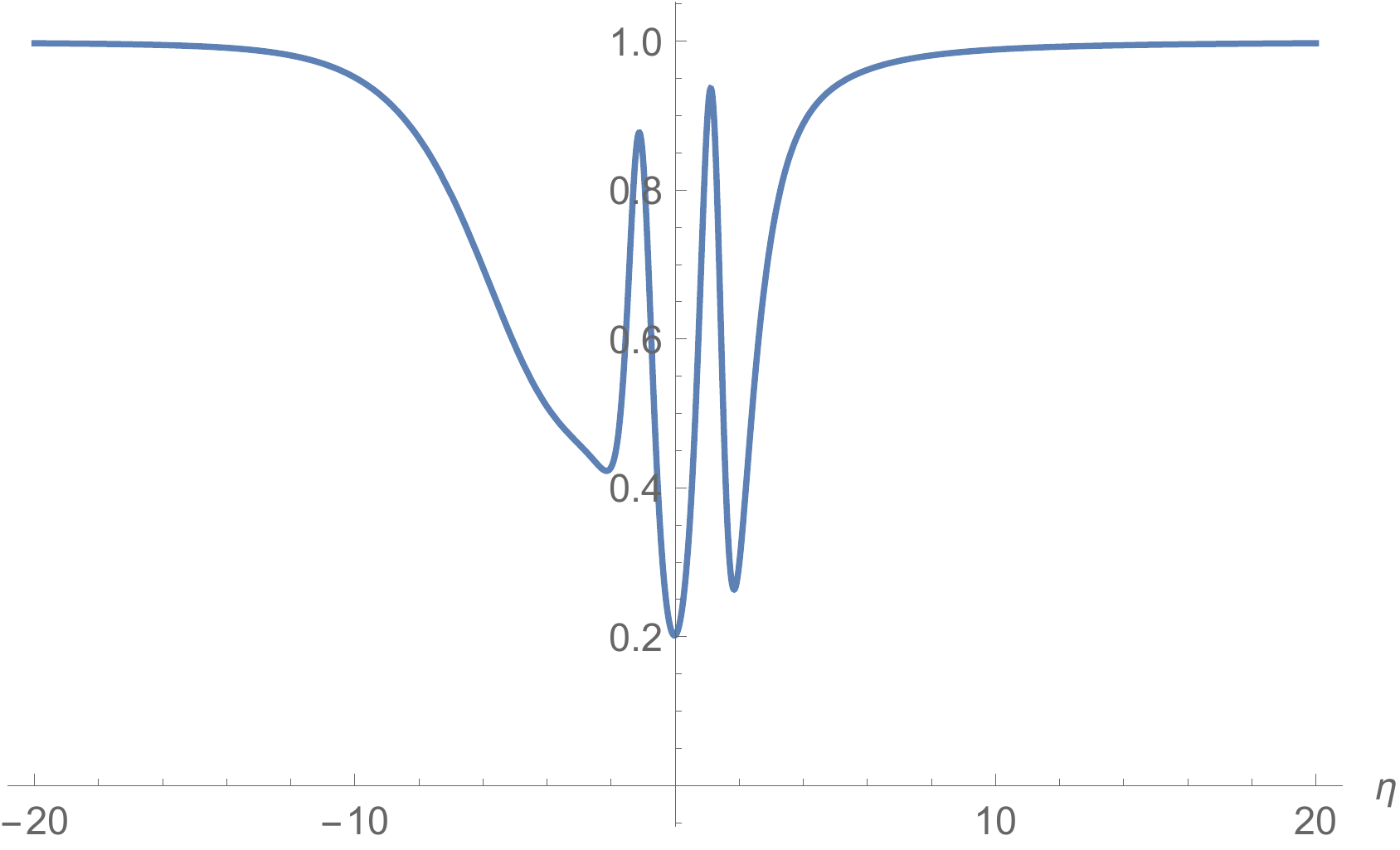}}
\caption{The model is ghost-free and gradient-stable since $U>0$
and $c_S^2>0$. During the expansion and contraction far
from the bounce phase, $c_S^2=1$.}\label{bh-model-cs}
\end{figure}

Therefor, with $c_1(\e)$, $c_2(\e)$ and $c_3(\e)$ satisfying
certain conditions, we will have a fully stable bounce model. As a
concrete example, setting
\begin{equation}\label{c123}
\bal
c_1(\e)&=k_1\left[1-\tanh\left(\frac{\e}{\tau_1}\right)\right],\\
c_2(t)&=\exp\left(-\frac{\e^2}{\tau^2_2}\right),\\
c_3(\e)&\equiv k_2, \\
\eal
\end{equation}
we plot Figs.\ref{bh-model-L} and \ref{bh-model-cs} with the
parameters $p_i=8$, $p_f=3$, $\tau_p=1$, $\e_p=0.7$ and $-\e_x=\tau_x=3$ in
(\ref{H bh}) and (\ref{n}), as well as $k_1=0.06$, $k_2=2$,
$\tau_1=2$ and $\tau^2_2=0.6$ in (\ref{c123}).
Fig.\ref{bh-model-L} shows that the coefficients $g_1(\e),
g_2(\e), g_3(\e)$ and $f_1(\e)$ in \eqref{free functions bh} have
been fixed. Fig.\ref{bh-model-cs} shows that the model is indeed
gradient-stable and ghost-free.

That the gravity should asymptotically approach GR requires $f_1\to
0$ in the asymptotic future. The asymptotic behavior of $f_1$ is
controlled by $c_2(\e)$. As a result, the sound speed squared
$c^2_S(+\infty)$ is (assume $c_1(+\infty)$ vanishes)
\[c^2_S(+\infty)=\frac{-xH'+Hx'}{3H^2x^2}. \]
Require $c_S^2(+\infty)=1$ and insert background \eqref{n}, one
finds
\begin{equation}\label{xf}
x_f=\frac{p_f}{3}.
\end{equation}
Similarly, $x_i$ is related to $p_i$ by requiring
$c_S^2(-\infty)=1$.
%Similarly, $x_i$ is related to $p_i$ through
%\begin{equation}\label{xi}
%c_S^2(-\infty)=\frac{-xH'+Hx'}{H^2x^2[3+4c_1(-\infty)x^2]}=1.
%\end{equation}

\subsection{In DHOST theory}\label{sec:dhost model}

The procedure is similar to that in Subsection \ref{sec:bh model}.
We write $P(N,\e)$, $Q(N,\e)$, $A(N,\e)$ and $B(N,\e)$ in
\eqref{ct1_DHOST_L_adm} as
\begin{equation}\label{free functions dhost}
\bal P(N,t)&=g_1(\e)\frac{1}{2N^2}+g_2(\e)\frac{1}{N^4}+g_3(\e),\\
Q(N,t)&=0,\\
A(N,t)&=\frac{1}{2}+\frac{g_4(\e)}{N^2},\\
B(N,t)&=b_0,\eal
\end{equation}
with $b_0\neq 0$ constant. So $\beta_1$, $\beta_2$, $\beta_3\neq
0$ in the quadratic order EFT of the DHOST theory \eqref{EFT}.

Substituting \eqref{free functions dhost} into \eqref{gamma}, we
have
\[\gamma\sim 2\mathcal{H}+b_0N',\]
Considering (\ref{n}), we have $N'(\pm\infty)=0$. This suggests
$\gamma(-\infty)\sim {\cal H}<0$ and $\gamma(+\infty)>0$. In other
words, the existence of $b_0$ only shifts the $\gamma$-crossing
point to $\e_0\neq 0$ instead of eliminating it. Therefore, the
condition \eqref{nsgl u} is still needed. To make $\mathcal{M}$
not divergent at $\e_0$, we might choose $g_4(\e)$ as
\begin{equation}\label{g4 dhost}
g_4=c_2(\e)\frac{ c_3(\e) \left(b_0 N'+2 \mathcal{H}\right)+N^2 (2
N-b_0)}{2 (b_0+2 N)},\qquad c_2(\e_0)=1.
\end{equation}
%where $\e_0$ is assumed to be the only zero of the denominator.
Thus with $c_1(\e)$ required in Eq.(\ref{nsgl u}), $c_2(\e)$ and
$c_3(\e)$ in Eq.(\ref{g4 dhost}), we could have a fully stable
bounce model based on the DHOST theory \eqref{ct1_DHOST_L_adm}.

\begin{figure}
    \centering
    \subfigure[$g_1(\e)$]{\includegraphics[width=2.5in]{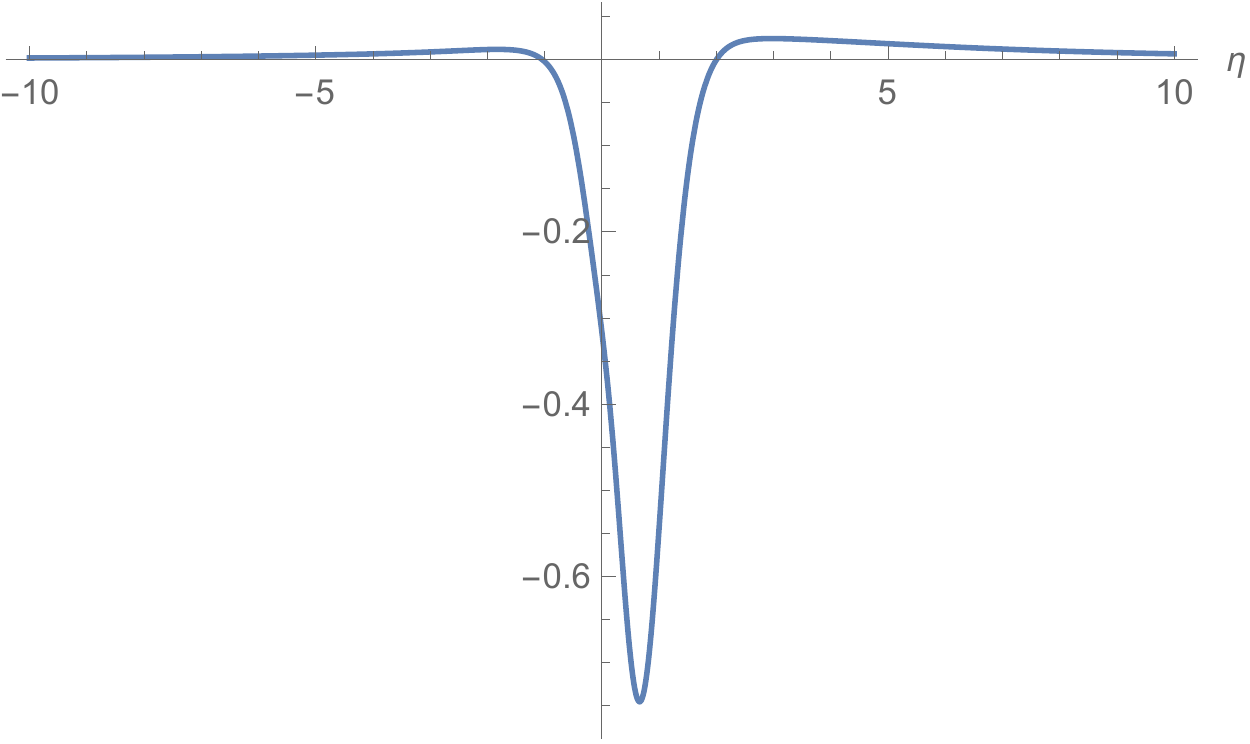}}
    \subfigure[$g_2(\e)$]{\includegraphics[width=2.5in]{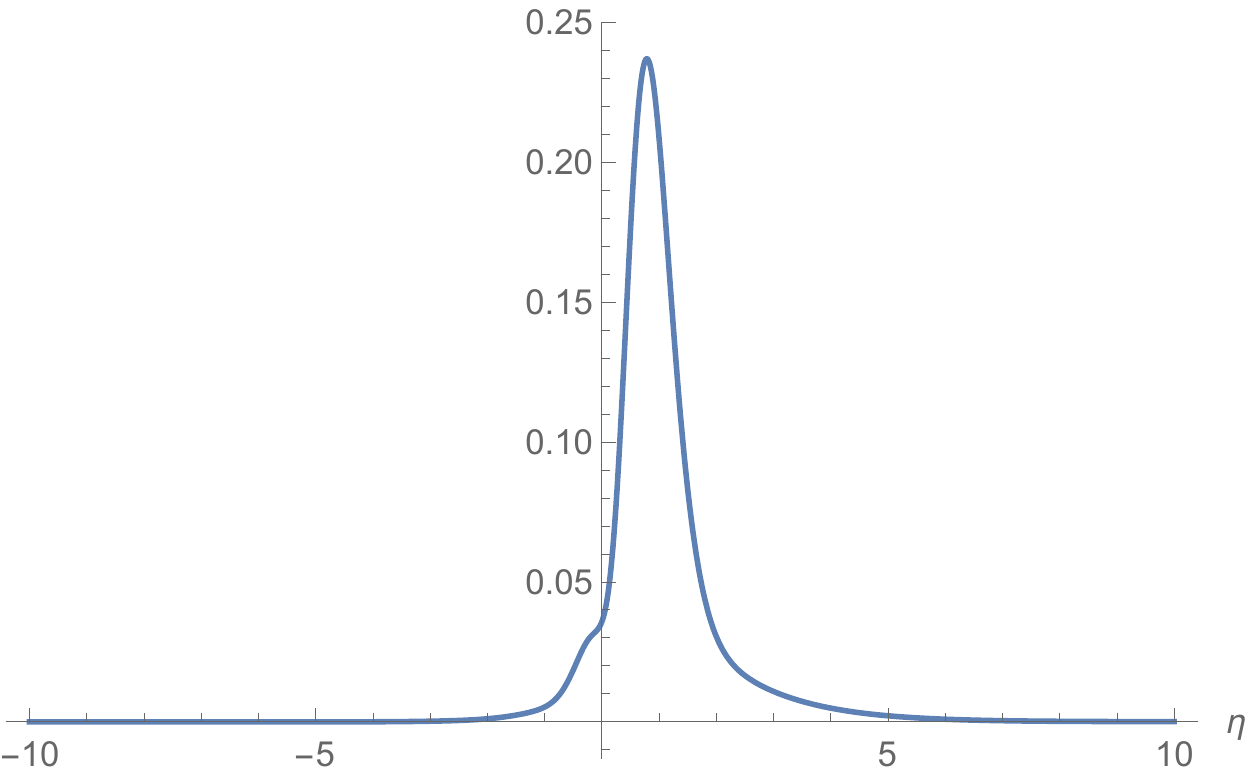}}
    \subfigure[$g_3(\e)$]{\includegraphics[width=2.5in]{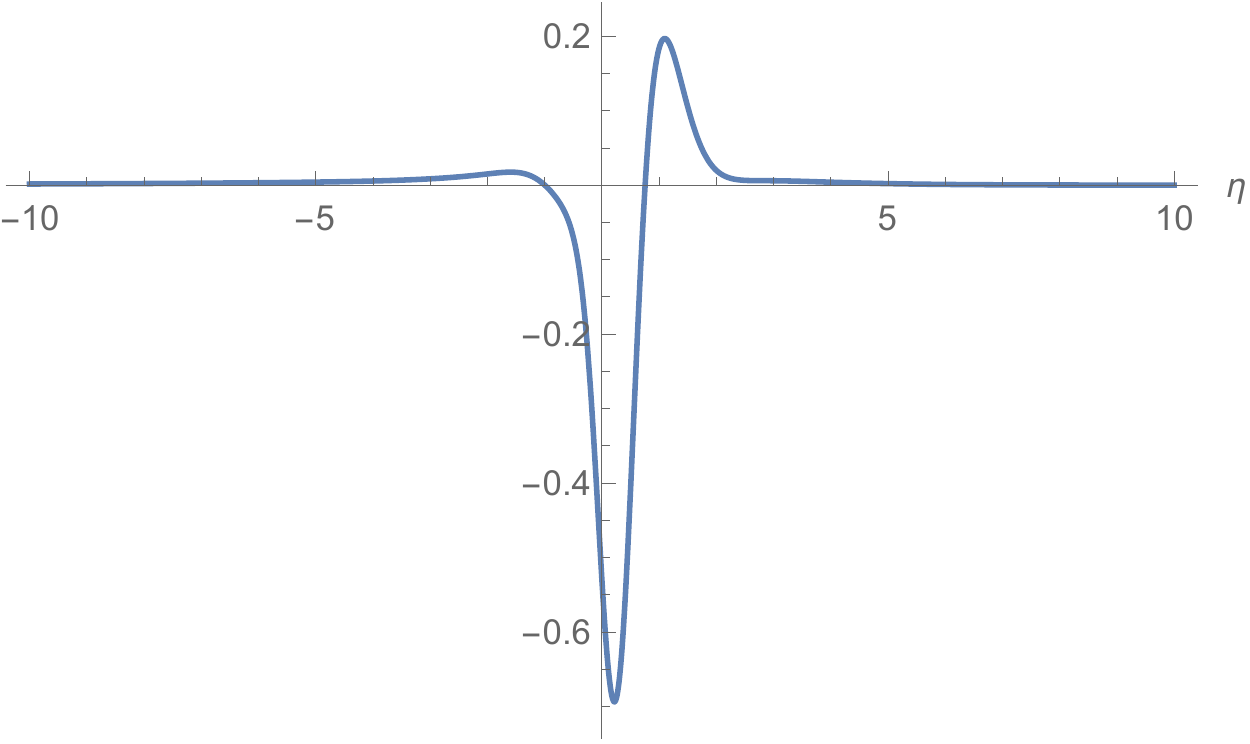}}
    \subfigure[$g_4(\e)$]{\includegraphics[width=2.5in]{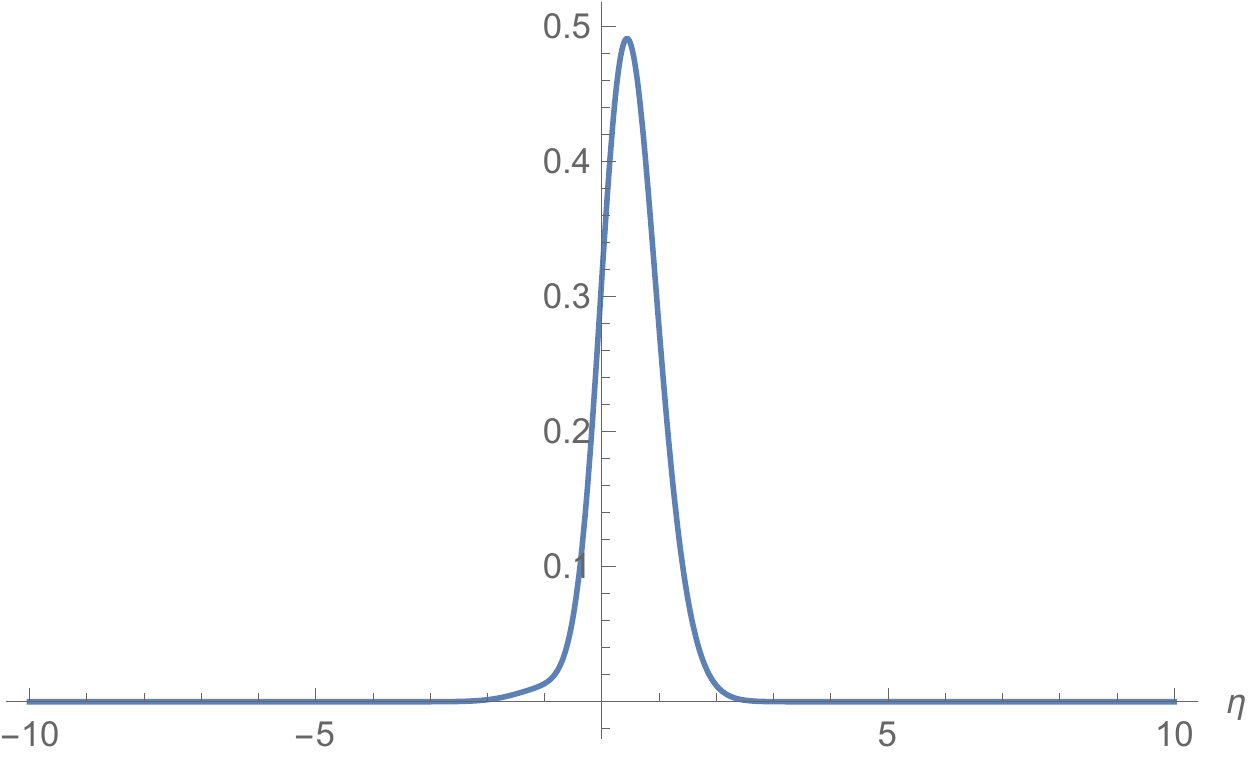}}
    \subfigure[$U$]{\includegraphics[width=2.5in]{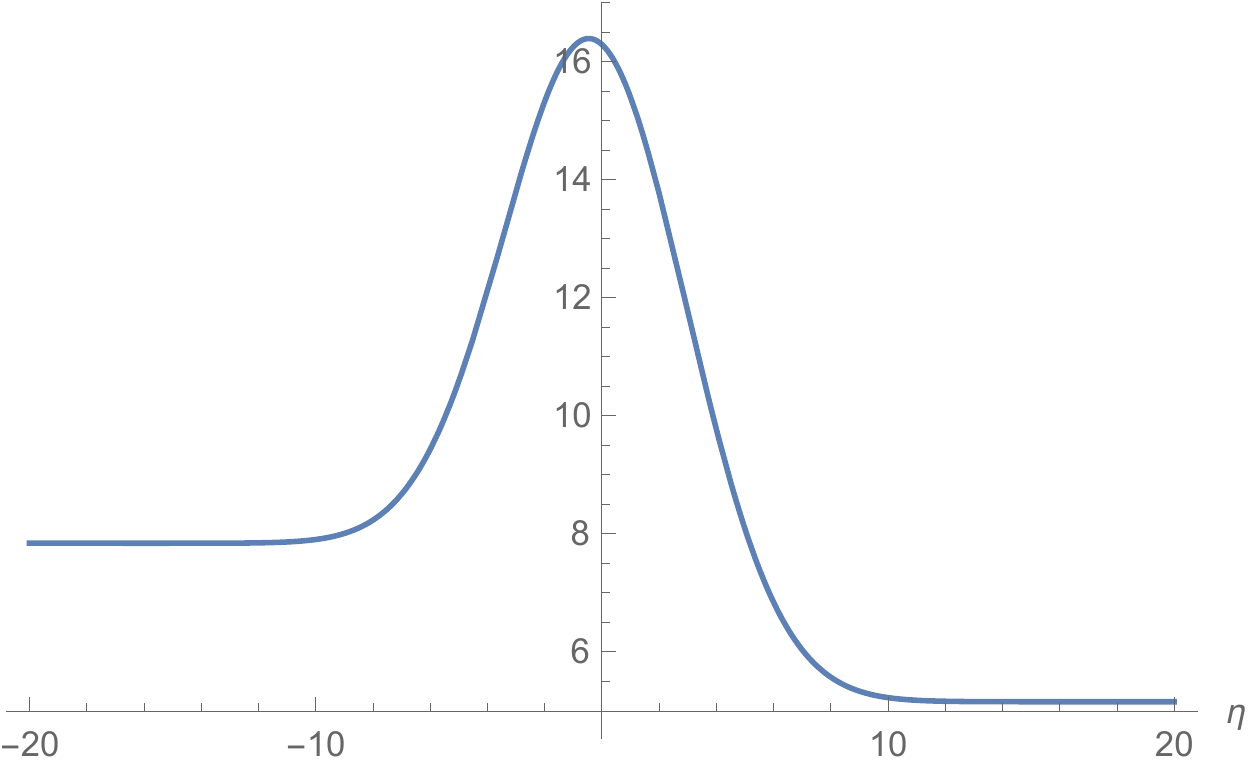}}
    \subfigure[$c_S^2$]{\includegraphics[width=2.5in]{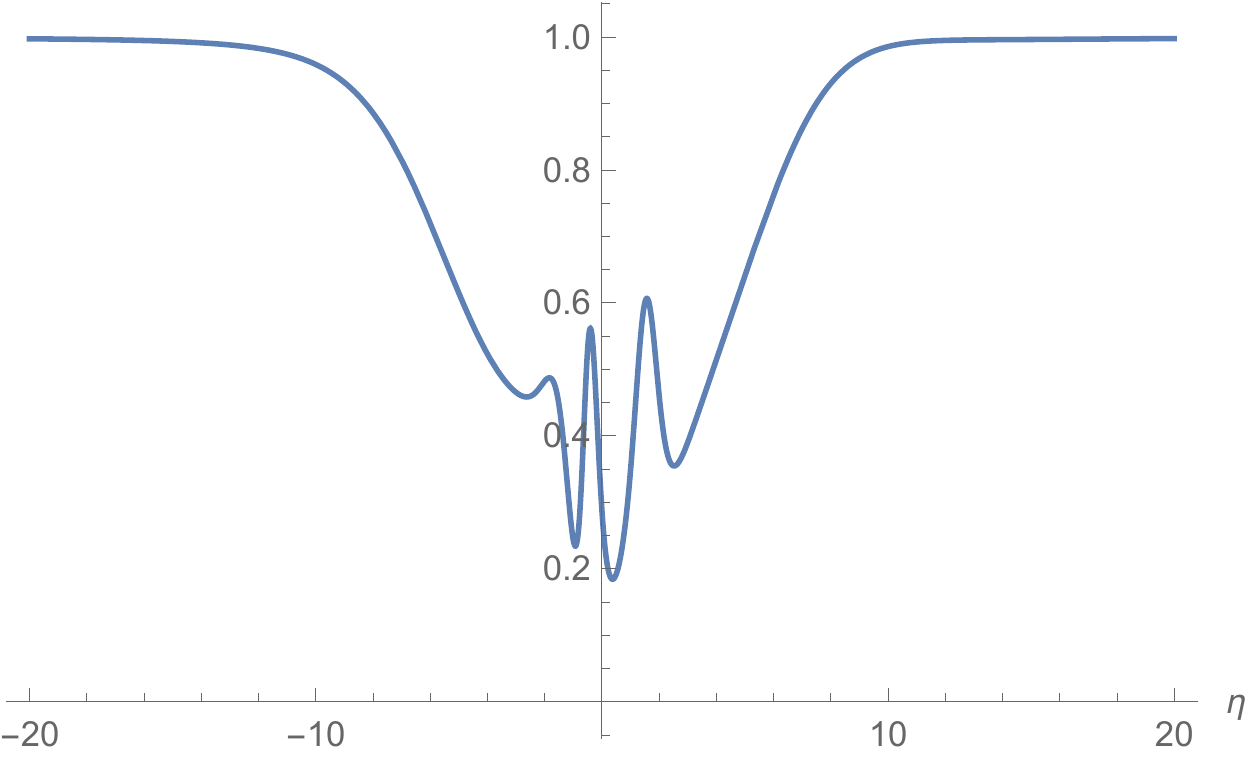}}
\caption{An example of the fully stable bounce model with a
constant DHOST term $b_0= 0.5$.}\label{dhost model 1}
\end{figure}

As a concrete example, setting
\begin{equation}\label{c123-2}
\bal
c_1(\e)&=k_1e^{-\e^2/\tau_1^2},\\
c_2(\e)&=e^{-(\e-\e_0)^2/\tau^2_2},\\
c_3(\e)& \equiv k_2, \\
\eal
\end{equation}
we plot Fig.\ref{dhost model 1} with the parameters $p_i=8$,
$p_f=3$, $\e_p=\e_x=0$, $\tau_p=1$ and $\tau_x=3$ in (\ref{H bh})
and (\ref{n}), as well as $\tau_1=20$ and $k_1=40$ in
(\ref{c123-2}), and $b_0=0.5$. Fig.\ref{dhost model 1} shows that
the model is indeed gradient-stable and ghost-free.

\section{Discussion}

We have constructed the spatially flat stable cosmological bounce
models with GR asymptotics in the $c_T=1$ beyond Horndeski theory
and in the full $c_T=1$ DHOST theory, respectively. In
Ref.\cite{Mironov:2018oec}, the stable bouncing solution with
$c_T=1$ has also been built in the beyond Horndeski theory (but
not in the full DHOST theory). Here, since we start straightly from
the Lagrangians with the constraint $c_T=1$, the procedure of
building models (even in full DHOST theory) is simpler.

It is well-known that the solutions of fully stable cosmological
bounce do exist in theories beyond Horndeski. Though the simplest
implementing is to work in the beyond Horndeski theory
\cite{Cai:2017dyi,Kolevatov:2017voe}, the stable bounce in a
full DHOST theory is still interesting for study, which might
bring unexpected results. In our implementing, we set the
parameter $B(\phi,X)=const.$ in the full $c_T=1$ DHOST theory
\eqref{ct1_DHOST_L_adm}, see \eqref{free functions dhost}.
Generally, it is not this case. The relevant issue will be studied
elsewhere.

The singularity of inflation implies that a bounce preceding
inflation might occur \cite{Piao:2003zm}, see also
\cite{Liu:2013kea,Liu:2013iha,Qiu:2015nha,Odintsov:2015zza,Mathew:2018rzn}.
Recently, it has been showed in Ref.\cite{Cai:2017pga} that the
bounce inflation scenario can explain the power deficit of CMB
TT-spectrum at low multipoles, specially the dip at multipole
$l\sim 20$. Thus it is interesting to embed the bounce models
built here into the corresponding scenario, which might bring
distinct imprint of DHOST terms in the CMB spectrum.

\textbf{Acknowledgments}

We thank Yong Cai for helpful discussions. This work is supported
by NSFC, Nos.11575188, 11690021.

\appendix
\section{On $g_1,g_2,g_3$}
We give the explicit algebraic solutions of $g_1,g_2,g_3$ here.
\subsection{The beyond Horndeski model}\label{apdx:bH}
Recall that $x\equiv1/N$ and $H=\h/N$.
\[\bal&\begin{split}
g_1=-\frac{1}{2 x^2}\big(&288 c_1 f_1^2 H^2 x^6+96 c_1 f_1 H^2
x^4+8 c_1 H^2 x^2+12 H x^3 f_1'\\&+36 f_1 H^2 x^2+12 f_1 x^3 H'+24
f_1 H x^2 x'+6 H^2+6 x H'\big)
\end{split}
\\&\begin{split}
g_2= -\frac{1}{8 x^4}\big(&-288 c_1 f_1^2 H^2 x^6-96 c_1 f_1 H^2
x^4-8 c_1 H^2 x^2-4 H x^3 f_1'\\&-60 f_1 H^2 x^2-4 f_1 x^3 H'-8
f_1 H x^2 x'-6 H^2-2 x H'\big)
\end{split}
\\&\begin{split}
g_3= \frac{1}{8} (&-18 H^2 + 8 c_1 H^2 x^2 - 36 f_1 H^2 x^2 + 96
c_1 f_1 H^2 x^4 + 288 c_1 f_1^2 H^2 x^6 \\&- 12 H x^3 f_1' - 6 x
H' - 12 f_1 x^3 H' - 24 f_1 H x^2 x')
\end{split}
\eal\]
\subsection{The DHOST model}\label{apdx:dhost}
\[\bal
g_1=&\frac{1}{8 N^3 \left(2 g_4+N^2\right)}\times\\
\big(&-36 b_0^2 c_1 g_4^2 N^3 \left(N'\right)^2-12 b_0^2 c_1 g_4
N^5 \left(N'\right)^2-b_0^2 c_1 N^7 \left(N'\right)^2+432 b_0^2
g_4^2 H N^2 N'\\&+144 b_0^2 g_4^2 N^2 N''-504 b_0^2 g_4^2 N
\left(N'\right)^2+72 b_0^2 N^4 g_4' N'+144 b_0^2 g_4 N^2 g_4'
N'\\&+360 b_0^2 g_4 H N^4 N'+120 b_0^2 g_4 N^4 N''-288 b_0^2 g_4
N^3 \left(N'\right)^2+72 b_0^2 H N^6 N'\\&+24 b_0^2 N^6 N''-18
b_0^2 N^5 \left(N'\right)^2-144 b_0 c_1 g_4^2 H N^3 N'-48 b_0 c_1
g_4 H N^5 N'\\&-4 b_0 c_1 H N^7 N'+720 b_0 g_4^2 H^2 N^2+96 b_0
g_4^2 N^2 H'-816 b_0 g_4^2 H N N'\\&-48 b_0 g_4^2 N N''+144 b_0
g_4^2 \left(N'\right)^2+120 b_0 H N^4 g_4'+240 b_0 g_4 H N^2
g_4'\\&-48 b_0 g_4 N g_4' N'-24 b_0 N^3 g_4' N'+576 b_0 g_4 H^2
N^4+120 b_0 g_4 N^4 H'-456 b_0 g_4 H N^3 N'\\&-48 b_0 g_4 N^3
N''+96 b_0 g_4 N^2 \left(N'\right)^2+108 b_0 H^2 N^6+36 b_0 N^6
H'-24 b_0 H N^5 N'\\&-12 b_0 N^5 N''+12 b_0 N^4
\left(N'\right)^2-144 c_1 g_4^2 H^2 N^3-48 c_1 g_4 H^2 N^5-4 c_1
H^2 N^7\\&-288 g_4^2 H^2 N-96 g_4^2 N H'+288 g_4^2 H N'-48 H N^3
g_4'-96 g_4 H N g_4'-192 g_4 H^2 N^3\\&-96 g_4 N^3 H'+192 g_4 H
N^2 N'-24 H^2 N^5-24 N^5 H'+24 H N^4 N'\big) \eal\]
\[\bal
g_2=&\frac{1}{32 N \left(2 g_4+N^2\right)}\times\\\big(&36 b_0^2
c_1 g_4^2 N^3 \left(N'\right)^2+12 b_0^2 c_1 g_4 N^5
\left(N'\right)^2+b_0^2 c_1 N^7 \left(N'\right)^2-288 b_0^2 g_4^2
H N^2 N'-96 b_0^2 g_4^2 N^2 N''\\&+456 b_0^2 g_4^2 N
\left(N'\right)^2-48 b_0^2 N^4 g_4' N'-96 b_0^2 g_4 N^2 g_4'
N'-216 b_0^2 g_4 H N^4 N'-72 b_0^2 g_4 N^4 N''\\&+264 b_0^2 g_4
N^3 \left(N'\right)^2-36 b_0^2 H N^6 N'-12 b_0^2 N^6 N''+18 b_0^2
N^5 \left(N'\right)^2+144 b_0 c_1 g_4^2 H N^3 N'\\&+48 b_0 c_1 g_4
H N^5 N'+4 b_0 c_1 H N^7 N'-432 b_0 g_4^2 H^2 N^2+816 b_0 g_4^2 H
N N'+16 b_0 g_4^2 N N''\\&-48 b_0 g_4^2 \left(N'\right)^2-72 b_0 H
N^4 g_4'-144 b_0 g_4 H N^2 g_4'+16 b_0 g_4 N g_4' N'+8 b_0 N^3
g_4' N'\\&-288 b_0 g_4 H^2 N^4-24 b_0 g_4 N^4 H'+456 b_0 g_4 H N^3
N'+16 b_0 g_4 N^3 N''-32 b_0 g_4 N^2 \left(N'\right)^2\\&-36 b_0
H^2 N^6-12 b_0 N^6 H'+24 b_0 H N^5 N'+4 b_0 N^5 N''-4 b_0 N^4
\left(N'\right)^2+144 c_1 g_4^2 H^2 N^3\\&+48 c_1 g_4 H^2 N^5+4
c_1 H^2 N^7+480 g_4^2 H^2 N+32 g_4^2 N H'-96 g_4^2 H N'+16 H N^3
g_4'\\&+32 g_4 H N g_4'+288 g_4 H^2 N^3+32 g_4 N^3 H'-64 g_4 H N^2
N'+24 H^2 N^5+8 N^5 H'\\&-8 H N^4 N'\big) \eal\]
\[\bal
g_3=&\frac{1}{32 N^5 \left(2 g_4+N^2\right)}\times\\\big(&36 b_0^2
c_1 g_4^2 N^3 \left(N'\right)^2+12 b_0^2 c_1 g_4 N^5
\left(N'\right)^2+b_0^2 c_1 N^7 \left(N'\right)^2-576 b_0^2 g_4^2
H N^2 N'-192 b_0^2 g_4^2 N^2 N''\\&+648 b_0^2 g_4^2 N
\left(N'\right)^2-96 b_0^2 N^4 g_4' N'-192 b_0^2 g_4 N^2 g_4'
N'-504 b_0^2 g_4 H N^4 N'-168 b_0^2 g_4 N^4 N''\\&+408 b_0^2 g_4
N^3 \left(N'\right)^2-108 b_0^2 H N^6 N'-36 b_0^2 N^6 N''+42 b_0^2
N^5 \left(N'\right)^2+144 b_0 c_1 g_4^2 H N^3 N'\\&+48 b_0 c_1 g_4
H N^5 N'+4 b_0 c_1 H N^7 N'-1008 b_0 g_4^2 H^2 N^2-192 b_0 g_4^2
N^2 H'+816 b_0 g_4^2 H N N'\\&-48 b_0 g_4^2 N N''+144 b_0 g_4^2
\left(N'\right)^2-168 b_0 H N^4 g_4'-336 b_0 g_4 H N^2 g_4'-48 b_0
g_4 N g_4' N'\\&-24 b_0 N^3 g_4' N'-864 b_0 g_4 H^2 N^4-216 b_0
g_4 N^4 H'+456 b_0 g_4 H N^3 N'-48 b_0 g_4 N^3 N''\\&+96 b_0 g_4
N^2 \left(N'\right)^2-180 b_0 H^2 N^6-60 b_0 N^6 H'+24 b_0 H N^5
N'-12 b_0 N^5 N''\\&+12 b_0 N^4 \left(N'\right)^2+144 c_1 g_4^2
H^2 N^3+48 c_1 g_4 H^2 N^5+4 c_1 H^2 N^7-288 g_4^2 H^2 N-96 g_4^2
N H'\\&+288 g_4^2 H N'-48 H N^3 g_4'-96 g_4 H N g_4'-288 g_4 H^2
N^3-96 g_4 N^3 H'+192 g_4 H N^2 N'\\&-72 H^2 N^5-24 N^5 H'+24 H
N^4 N'\big) \eal\]

\end{document}